\documentclass[prl,amsmath,twocolumn,amssymb,superscriptaddress]{revtex4-1}
\usepackage{times}
\usepackage{graphicx}
\usepackage{amsmath, braket}
\usepackage{amssymb}
\usepackage{natbib}
\usepackage[usenames]{xcolor}
\usepackage{mhchem}
\usepackage{siunitx} 
\usepackage{xspace} 
\usepackage[colorlinks]{hyperref}

\newcommand{\tas}{\ce{TaS2}\xspace}
\newcommand{\nbse}{\ce{NbSe2}\xspace}
\newcommand{\hBN}{h-\ce{BN}\xspace}

\newcommand{\Kval}{$\mathbf{K}$\xspace}
\newcommand{\Kpval}{$\mathbf{K'}$\xspace}
\newcommand{\K}{K\xspace}
\newcommand{\Kp}{K$^\prime$\xspace}

\newcommand{\Rn}{\ensuremath{R_n}}
\newcommand{\Rgr}{\ensuremath{R_\mathrm{graphite}}}
\newcommand{\muB}{\ensuremath{\mu_\mathrm{B}}}
\newcommand{\kB}{\ensuremath{k_\mathrm{B}}}
\newcommand{\kF}{\ensuremath{k_\mathrm{F}}}

\newcommand{\Tc}{\ensuremath{T_c}}
\newcommand{\Tcz}{\ensuremath{T_{c0}}}

\newcommand{\Hp}{\ensuremath{H_\mathrm{P}}}
\newcommand{\Hso}{\ensuremath{H_\mathrm{so}}}
\newcommand{\Hperp}{\ensuremath{H_\perp}}
\newcommand{\Hpara}{\ensuremath{H_\parallel}}
\newcommand{\Hc}{\ensuremath{H_{c2}^\parallel}}

\newcommand{\Bpara}{\ensuremath{B_\parallel}}
\newcommand{\Bso}{\ensuremath{B_\mathrm{so}}}
\newcommand{\Bsov}{\ensuremath{\mathbf{B}_\mathrm{so}}}

\newcommand{\Dso}{\ensuremath{\Delta_\mathrm{so}}}
\newcommand{\Dvb}{\ensuremath{\Delta_\mathrm{VB}}}

\newcommand{\alphaR}{\ensuremath{\alpha_\mathrm{R}}}

\begin{document}

\title{Tuning Ising superconductivity with layer and spin-orbit coupling \\in two-dimensional transition-metal dichalcogenides}

\author{Sergio~C.~de~la~Barrera} \email{sergio@phys.cmu.edu}
\author{Michael~R.~Sinko}
\author{Devashish~P.~Gopalan}
\affiliation{Department of Physics, Carnegie Mellon University, Pittsburgh, PA 15213}
\author{Nikhil~Sivadas}
\affiliation{Department of Physics, Carnegie Mellon University, Pittsburgh, PA 15213}
\affiliation{School of Applied and Engineering Physics, Cornell University, Ithaca, NY 14853}
\author{Kyle~L.~Seyler}
\affiliation{Department of Physics, University of Washington, Seattle, WA 98195}
\author{Kenji~Watanabe}
\author{Takashi~Taniguchi}
\affiliation{Advanced Materials Laboratory, National Institute for Materials Science, Tsukuba, Ibaraki 305-0044, Japan}
\author{Adam~W.~Tsen}
\affiliation{Institute for Quantum Computing and Department of Chemistry, University of Waterloo, Waterloo, Ontario N2L 3G1, Canada}
\author{Xiaodong~Xu}
\affiliation{Department of Physics, University of Washington, Seattle, WA 98195}
\affiliation{Department of Materials Science and Engineering, University of Washington, Seattle, WA 98195}
\author{Di~Xiao}
\author{Benjamin~M.~Hunt} \email{bmhunt@andrew.cmu.edu}
\affiliation{Department of Physics, Carnegie Mellon University, Pittsburgh, PA 15213}

\begin{abstract}
Systems that simultaneously exhibit superconductivity and spin-orbit coupling are predicted to provide a route toward topological superconductivity and unconventional electron pairing, driving significant contemporary interest in these materials.
Monolayer transition-metal dichalcogenide (TMD) superconductors in particular lack inversion symmetry, enforcing a spin-triplet component of the superconducting wavefunction that increases with the strength of spin-orbit coupling. 
In this work, we present an experimental and theoretical study of two intrinsic TMD superconductors with large spin-orbit coupling in the atomic layer limit, metallic 2H-\tas and 2H-\nbse.
For the first time in \tas, we investigate the superconducting properties as the material is reduced to a monolayer and show that high-field measurements point to the largest upper critical field thus reported for an intrinsic TMD superconductor.
In few-layer samples, we find that the enhancement of the upper critical field is sustained by the dominance of spin-orbit coupling over weak interlayer coupling, providing additional platforms for unconventional superconducting states in two dimensions.
\end{abstract}
\maketitle

\begin{figure}[ht!]
  \begin{center}
\includegraphics[width=85 mm]{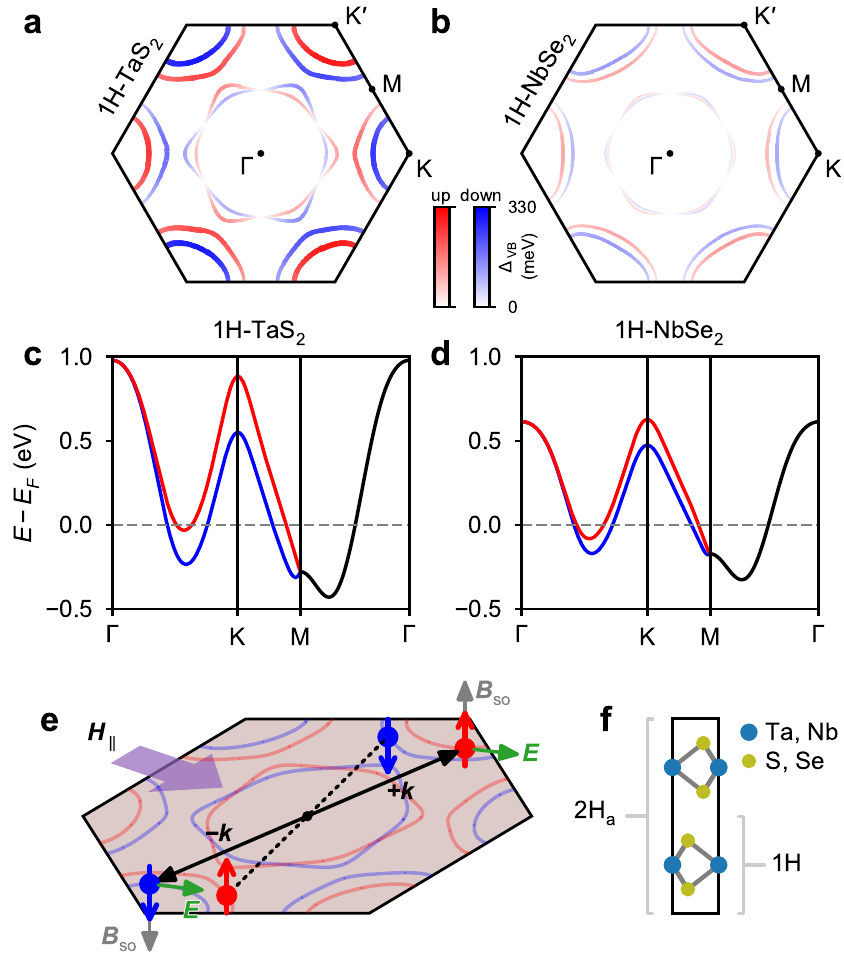} 
\caption{ \textbf{Electronic structure of monolayer metallic transition-metal dichalcogenides.}
(a) Spin-projected Fermi surface of monolayer \tas and (b) \nbse computed by density functional theory (DFT).
Red corresponds to one $S_z$ projection and blue to the opposite (e.g. up and down, respectively).
Variation in the shading and curve thickness indicates the magnitude of spin-splitting in the valence band $\Dvb$ due to spin-orbit coupling, with the color scale being shared between (a) and (b) to emphasize the difference in magnitudes.
(c) Relevant bands around the Fermi level for monolayer \tas and (d) \nbse from DFT, with spin polarization corresponding to colors in (a), black bands being spin-degenerate.
(e) Schematic of a Cooper pair with spins pinned to the out-of-plane direction due to effective field $\Bsov \propto \mathbf{E \times k}$ resulting from planar crystal field and momentum.
(f) Crystal structure of 2H$_a$-\ce{MX2} (with transition metal atoms directly above one another along the c-axis), viewed along [100] direction for $\mathrm{M}\in\{\mathrm{Nb, Ta}\}$ and $\mathrm{X}\in\{\mathrm{S, Se}\}$ with 1H (monolayer) substructure indicated.}
\label{fig-bands}
  \end{center}
\end{figure}

Cooper pairing in type-II superconductors is typically destroyed by external magnetic fields due to coupling between the applied field and electron orbital and spin degrees of freedom.
For fields applied in the plane of sufficiently thin superconductors, orbital effects are suppressed due to the reduced dimensionality, providing some protection for superconductivity at enhanced fields.
In this limit, the dominant mechanism for breaking superconducting order is Pauli paramagnetism, in which the upper critical field $\Hc$ is given by the Chandrasekhar-Clogston (or Pauli) limit of $\Hc = \Hp \equiv \beta\Tcz$ at $T=0$, with $\beta=\SI{1.84}{T/K}$ and superconducting transition temperature $\Tcz$ \cite{clogston_upper_1962,chandrasekhar_note_1962}.
However, recent measurements have shown that superconductivity in intrinsic (i.e. metallic) and electrolyte-gated monolayer TMDs survives in the presence of in-plane fields significantly beyond the Pauli limit due to a mechanism known as Ising pairing \cite{lu_evidence_2015,xi_ising_2016,saito_superconductivity_2016}, in which a particular type of Dresselhaus spin-orbit coupling (SOC) pins the Cooper pair spins to the out-of-plane direction \cite{xiao_coupled_2012,hsu_topological_2017}, termed Ising SOC.
In the 1H-phase of monolayer TMDs (see Fig.~\ref{fig-bands}f), the out-of-plane mirror symmetry forces the crystal electric field $\mathbf{E}$ to be in-plane.
Thus, for electron motion in the same $x$-$y$ plane, intrinsic SOC gives rise to an effective magnetic field $\Bsov \propto \mathbf{E \times k}$ that is directed out-of-plane, leading to a momentum-dependent energy splitting between the spin states $g \muB \Bso(\mathbf{k})$ that changes sign upon inversion through the Brillouin zone center.

This spin splitting naturally leads to Cooper pairing between an electron in one of the two spin-split Fermi surfaces around \Kval with its time-reversed pair, of opposite spin and momentum, in \Kpval.
The two Fermi surfaces give rise to two distinct populations of Cooper pairs, one each from the upper and lower spin-split bands (see Fig.~\ref{fig-bands}a--d), with differing densities of states at the Fermi level.
The superconducting state built up of these Cooper pairs is now known as Ising superconductivity, evocative of the notion that the constituent electrons have one of two spin projections in the out-of-plane direction.

In atomically-thin TMD superconductors, there are many other effects that can complicate this basic picture of Ising superconductivity in monolayers:  (i) Additional Cooper pair channels allowed by the band structure beyond \K and \Kp pairing, (ii) coupling between the layers in few-layer samples, (iii) Rashba SOC, which competes against Ising SOC to tilt the electron spins in-plane \cite{lu_evidence_2015,saito_superconductivity_2016,lu_full_2017}, and (iv) extrinsic effects such as intervalley scattering \cite{ilic_enhancement_2017}. The relative importance of these effects in modifying the Ising protection of $\Hc$ is an open experimental and theoretical question.    

In this work, we study 2H$_a$-\tas, an intrinsic TMD superconductor with the same crystal symmetry and similar electronic structure as \nbse, but with stronger SOC.
Experimentally, we compare the superconducting properties of atomically thin 2H$_a$-\tas, with a large atomic SOC contribution from the heavy \ce{Ta} atoms, with those of 2H$_a$-\nbse (hereafter \tas and \nbse, respectively).
We isolate ultrathin \tas to the monolayer limit, confirming for the first time that there is in fact a stable 1H polytype with a superconducting phase, and extend existing measurements \cite{navarro-moratalla_enhanced_2016} of $\Tcz$ as a function of the number of layers $N$ down to the monolayer limit.
We show that the upper critical field $\Hc(T)$ is significantly enhanced in monolayer \tas relative to \nbse, compelling evidence of the Ising SOC origin of pairing protection in these intrinsic metallic TMDs.
We perform first-principles calculations of the band structures and Fermi surfaces of monolayer \tas and \nbse, including spin-orbit coupling, and we analyze the bands to quantify the role of additional Cooper pairing, previously neglected, from around the $\Gamma$ pocket.
We measure $\Hc(T)$ in several few-layer devices of \tas and \nbse and observe a large enhancement of $\Hc$ above $\Hp$ in 2L and 4L devices which is close to that of 3L and 5L devices, despite the restoration of inversion symmetry in the even-layer-number devices.
To provide insight into this persistent enhancement of $\Hc$, we calculate the interlayer coupling energy $t_\perp$ for \nbse and \tas and show that the trend of $\Hc$ as a function of the number of layers, $N$, follows the ratio of the interlayer coupling energy and SOC strength $t_\perp(N)/\Dso$.
Finally, we measure $\Hc(T)$ in few-layer \nbse and \tas in a crucial low-temperature regime, down to \SI{300}{mK}, where differences among the various theoretical models become evident \cite{lu_evidence_2015,saito_superconductivity_2016,xi_ising_2016,wakatsuki_proximity_2016,nakamura_odd-parity_2017,ilic_enhancement_2017}.

\section{Results}
\begin{figure}[ht!]
	\begin{center}
\includegraphics[width=89 mm]{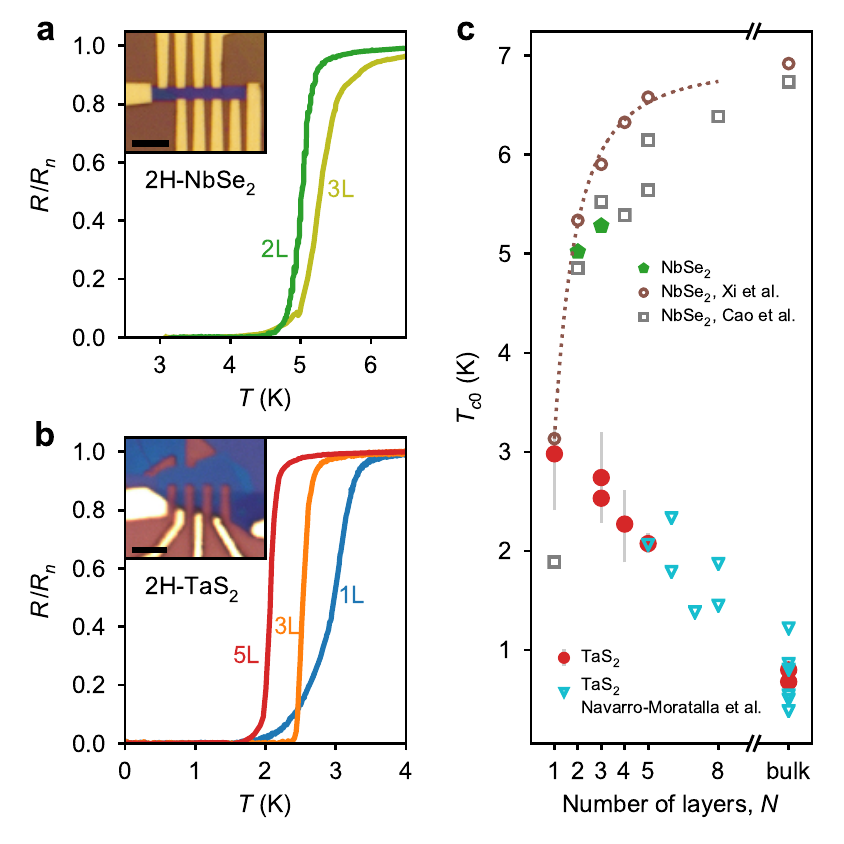}
\caption{ \textbf{Superconductivity in zero magnetic field of atomically-thin \tas and NbSe$_2$.}
(a) Temperature dependence of the normalized longitudinal resistance for bilayer (2L) and trilayer (3L) \nbse around the superconducting transition.
Optical image of the 3L sample shown in inset.
(b) Resistance of monolayer (1L), trilayer (3L), and five-layer (5L) \tas samples, exhibiting a reverse trend in the $\Tcz$ with decreasing thickness.
Optical image of 3L sample shown in inset.
Scale bars are \SI{4}{\micro\meter}.
(c) Compilation of superconducting transition temperatures $\Tcz$ as a function of thickness for \tas and \nbse samples from this work as well as from \citet{navarro-moratalla_enhanced_2016,xi_ising_2016,cao_quality_2015}, using $50\%$ of normal state resistance to define the transitions.
Error bars on \tas data denote temperature at $10\%$ and $90\%$ of normal state resistances.
Dotted line follows the fitting curve used in Ref.~\citenum{xi_ising_2016} for \nbse.
}
		\label{fig-zerofield}
	\end{center}
\end{figure}

We fabricated several multiterminal transport devices from \tas and \nbse exfoliated from bulk crystals, capped with boron nitride (BN) inside a nitrogen-filled glove box, and contacted with graphite in series with \ce{Cr/Pd/Au} leads (more details are in the Methods). 
Figure~\ref{fig-zerofield} shows a measurement of the longitudinal resistance $R_{xx}(T)$ of five samples in our study: bilayer (2L) and trilayer (3L) \nbse and of monolayer (1L), trilayer (3L), and five-layer (5L)  \tas.
All samples show a transition from the normal state (with resistance $\Rn$ of the order of \SI{100}{\ohm} per square for all samples) to a zero-resistance state at a temperature $\Tcz$, which we take by convention to be defined by $R(\Tcz)=0.5\Rn$.
For $T>\Tcz$, a rounding of the transition is observed that is similar in all of our samples.
This is indicative of the enhanced fluctuations in two dimensions and can be described by fitting, for example, the Aslamazov-Larkin or the Halperin-Nelson formulae to these $R(T)$ curves for $T>\Tcz$ \cite{hsu_superconducting_1992}.
For $T<\Tcz$, a finite-resistance tail develops with a degree of rounding that varies from sample to sample.
In Fig.~\ref{fig-zerofield}b this is seen clearly if one compares the 1L and 3L data.
We ascribe this behavior to effects of the finite size of our samples \cite{hsu_superconducting_1992}.

\tas is known to exhibit a surprising trend in the superconducting critical temperature $\Tc$ as a function of thickness \cite{navarro-moratalla_enhanced_2016}; whereas $\Tcz$ decreases as the number of layers is reduced for \nbse, in \tas the opposite trend is observed down to 5 layers, the thinnest sample previously reported.
Here, we show that this striking trend continues to the monolayer limit (Fig.~\ref{fig-zerofield}c), however, the mechanism behind this enhancement of $\Tcz$ is a subject of ongoing debate \cite{talantsev_origin_2017}.

For a given layer number $N$, there is significant amount of scatter in the $\Tcz$ data for both \tas and \nbse.
For example, for bilayer \nbse, measurements of $\Tcz$ from this work and from Refs.~\citenum{xi_ising_2016} and \citenum{cao_quality_2015} span the range from \SIrange{4.9}{5.3}{K}.
This variation within a given $N$ may be due to effects from the substrate or to varying amounts of disorder from sample to sample.  An effect that can have a much larger impact, particularly in the case of \tas, is intercalation by organic and non-organic molecules \cite{klemm_pristine_2015}. To exclude the possibility of unintentional intercalation of the \tas crystals, we performed control experiments on bulk devices fabricated alongside the 1L and 3L samples, subject to the same fabrication processes (see SI for details).


To obtain the cleanest results possible, the data shown in Fig.~\ref{fig-zerofield} were taken within a few days of exfoliation of each crystal.
However, despite the \hBN encapsulation intended to protect the \tas crystals during the brief periods of ambient exposure between experiments, we did observe noticeable degradation in the superconducting properties within a few weeks to a few months of the devices being fabricated for all of the devices.
For the 1L device in particular, we found that the monolayer portion of the device degraded away entirely over a period of two months, leaving open the possibility that an even cleaner 1L sample might exhibit even more pronounced enhancement than what we report here.

	\begin{figure*}[ht!]
	\begin{center}
\includegraphics[width=178 mm]{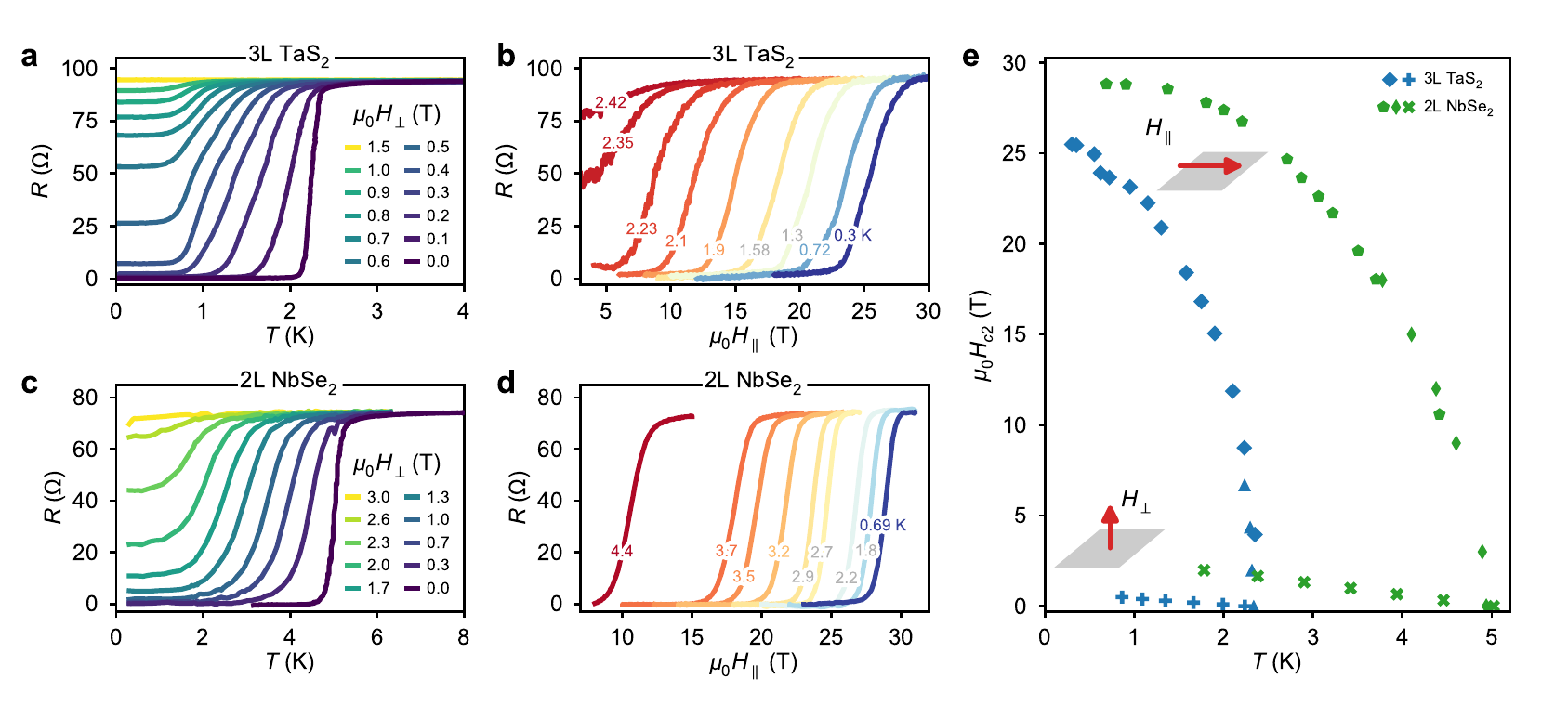}
\caption{ \textbf{Perpendicular and parallel magnetic field dependence.}
(a) Temperature dependence of longitudinal resistance of trilayer (3L) \tas in the presence of an applied magnetic field $\Hperp$ in the out-of-plane direction.
(b) Magnetic field dependence of the \tas resistance for a field $\Hpara$ applied in an in-plane direction, for a few constant temperatures as indicated.
The field value $H_{c2}^\parallel$ at which the resistance transitions to a zero-resistance state at a fixed temperature is equivalent to the transition temperature $\Tc$ of the superconducting state for a fixed field.
(c) Temperature dependence of the bilayer (2L) \nbse sample for a few perpendicular fields.
(d) In-plane field dependence of the same \nbse sample for a range of constant temperatures.
(e) Temperature dependence of the upper critical field $H_{c2}$ of both samples as extracted from the $50\%$ normal state resistances from the data shown in (a)--(d).}
		\label{fig-perpandparallelfield}
	\end{center}
\end{figure*}

We turn now to our investigation of atomically-thin TMD superconductors in the presence of magnetic fields perpendicular and parallel to the 2D plane.
Figure~\ref{fig-perpandparallelfield} shows the behavior of representative devices of \tas and \nbse.
In perpendicular field, superconductivity is destroyed when the total area occupied by the normal cores of vortices is comparable to the total area of the sample, as in three-dimensional (type-II) superconductors.
This leads to the Ginzburg-Landau expression for the upper critical field, $H_{c2}^\perp(T)=\frac{\Phi_0}{2\pi\xi_\mathrm{GL}(0)^2}(1-T/\Tcz)$, which allows us to estimate the coherence length $\xi_\mathrm{GL}(0) \approx \SI{20}{nm}$ for the 3L~\tas and \SI{10}{nm} for the 2L~\nbse.
In both of these samples, and indeed in all of the devices that we have studied, at finite perpendicular fields less than $H_{c2}^\perp$, the resistance of the devices does not go to zero as $T\rightarrow0$ but rather saturates to a finite value (Fig.~\ref{fig-perpandparallelfield}a and Fig.~\ref{fig-perpandparallelfield}c).
The nature of this finite-conductivity state at $T=0$ and $\Hperp\neq0$ has been discussed in Refs.~\citenum{tsen_nature_2016} and \citenum{saito_metallic_2015} and further discussion will be deferred to a future work, but we note it here to distinguish the zero-temperature behavior in perpendicular field from that in parallel field.

In Fig.~\ref{fig-perpandparallelfield}b and Fig.~\ref{fig-perpandparallelfield}d, we show the dependence of the resistance of the same \tas and \nbse devices as the parallel magnetic field (in the plane of the 2D crystals) is varied at fixed temperatures.  For some devices we also perform the measurement of $\Hc$ by fixing the parallel field and sweeping the temperature (see Methods).  At the lowest temperatures, superconductivity in the atomically-thin crystals survives up to very large parallel magnetic fields: \SI{25}{T} for 3L~\tas and \SI{28}{T} for 2L~\nbse, corresponding to an anisotropic enhancement $\Hc/H_{c2}^\perp$ of $32\times$ and $8\times$ the upper critical fields in the perpendicular orientation, respectively.
The anisotropy is even larger for monolayer TMDs, as will be discussed next.

\begin{figure*}[ht!]
	\begin{center}
\includegraphics[width=178 mm]{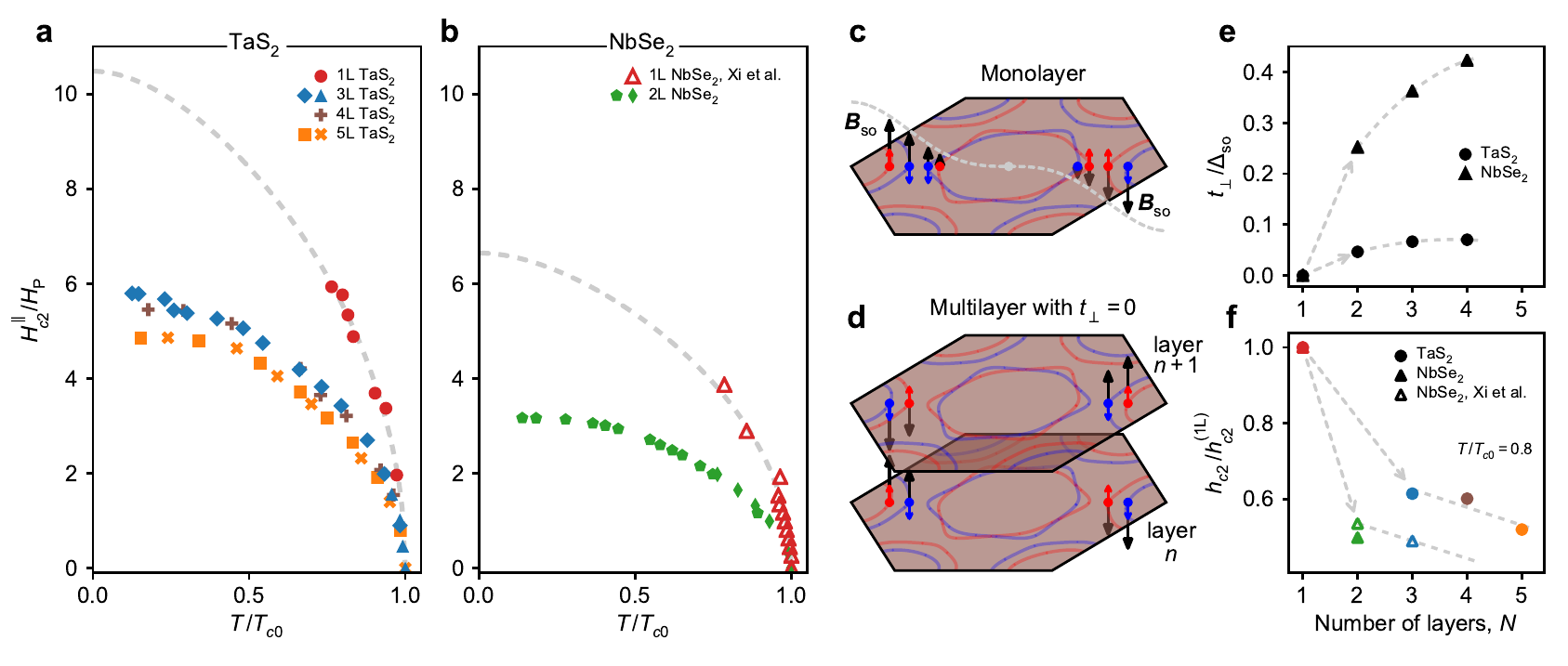}
\caption{ \textbf{Ising superconductivity in single-layer and few-layer \tas and \nbse.}
(a) Parallel upper critical field normalized to Pauli paramagnetic limit, $H_{c2}^\parallel/\Hp$, as a function of reduced temperature $T/\Tcz$ for monolayer (1L), trilayer (3L), four-layer (4L), and five-layer (5L) \tas samples, and
(b) monolayer (1L, from \citet{xi_ising_2016}) and bilayer (2L) \nbse.
Dashed lines follow pair-breaking fit to 1L data.
(c) Illustration of spin states along intersection Fermi surface and line of high symmetry from 1L~\tas,  experiencing varying $\Bsov(\mathbf{k})$ throughout Brillouin zone.
Spin projections indicated by color, as in Fig.~\ref{fig-bands}.
Dashed line marks the envelope of $\Bsov$ along \Kp--$\Gamma$--\K.
(d) Schematic Fermi surface of multilayer \tas ignoring interlayer coupling, $t_\perp$, showing opposite $\Bsov$ in neighboring layers due to inversion of crystal field between layers.
(e) Ratio of interlayer coupling energies to spin-orbit coupling strength $t_\perp/\Dso$ as a function of the number of layers for \tas and \nbse extracted from DFT, with $\Dso^{\tas}=\SI{156}{meV}$ and $\Dso^{\nbse}=\SI{62.6}{meV}$.
Dashed lines are provided as guides to the eye.
(f) Dependence of the reduced upper critical field $h_{c2} \equiv \Hc/\Hp$ evaluated at $T/\Tcz=0.8$ on the number of \tas or \nbse layers, including \nbse data from \citet{xi_ising_2016}.
Data are normalized by the reduced upper critical field of a monolayer $h_{c2}^{(\mathrm{1L})}$ to enable direct comparison.
}
		\label{fig-alldata}
	\end{center}
\end{figure*}

In monolayer \tas, we find that for $T<\SI{2}{K}$, the upper critical field in the parallel orientation is larger than the highest field available (\SI{34.5}{T}; Fig.~\ref{fig-alldata}a) in the experimental apparatus, whereas in a perpendicular field the superconductivity is quenched at a field near \SI{1.2}{T} as $T\rightarrow 0$.
The qualitative behavior of monolayer \nbse is similar \cite{xi_ising_2016}, but with slightly modified temperature and field scales.
To facilitate a quantitative comparison between the two materials, we plot the in-plane upper critical field $\Hc(T)$ normalized to the Pauli limit $\Hp$ versus the reduced temperature $T/\Tcz$.
Figure~\ref{fig-alldata} shows a summary of our $\Hc(T)$ data for 1L, 3L, 4L, and 5L~\tas (Fig.~\ref{fig-alldata}a), along with 2L~\nbse superimposed with 1L-\nbse data (Fig.~\ref{fig-alldata}b) from Ref.~\citenum{xi_ising_2016}.
On this scale, it is clear that these materials continue to superconduct well above the Pauli limit, $\Hp$, and that the slope of the phase boundary $d\Hc/dT$ between normal and superconducting states near $T/\Tcz\rightarrow 1$ is strikingly steeper for the monolayer samples compared to the few-layer ones.

\section{Discussion}
In the absence of orbital effects and spin-orbit coupling, the upper critical field $\Hc$ is determined by comparing the superconducting condensation energy $E_c = \frac{1}{2}N(0)\Delta_0^2$ with the Pauli paramagnetic energy $E_P = \frac{1}{2}\chi_n (\Hpara)^2$, which coincide when $ \Hpara = \Hc$ ($\equiv \Hp$, the Pauli limit).  Here, $\Delta_0$ is the zero-temperature superconducting gap and $N(0)$ is the density of states at the Fermi level.  The Pauli paramagnetic susceptibility of free electrons, $\chi_n=2\muB^2 N(0)$, can be significantly modified in the presence of the SOC field $\Hso=|\Bsov|/\mu_0$.
In the limit of $\Hso \gg \Hpara $, the in-plane magnetic moment operator ($\propto \sigma_x$) can be treated as a perturbation to the spin-orbit coupling ($\propto \tau \sigma_z$).
The in-plane magnetic moment is then zero in leading order but equal to $\mu_\parallel = \muB (\Hpara/\Hso)$ in second order; the spin splitting is correspondingly increased by an amount $\muB \Hpara^2/\Hso$.  
(This second-order perturbative determination of $\mu_\parallel$ is similar to van Vleck paramagnetism \cite{kittel_introduction_2005}.)
It is this reduced in-plane magnetic moment $\mu_\parallel \ll \muB$ that is responsible for the large increase in $\Hc$ above $\Hp$. 

We now consider quantitatively the effect of the modified in-plane magnetic moment on $\Hc$.
The in-plane field is a magnetic perturbation that breaks time-reversal symmetry (unlike $\Hso$) and so
we employ the standard pair-breaking theory using a pair-breaking energy $\alpha = \muB \Hpara^2/\Hso$ \cite{maki_gapless_1969,tinkham_introduction_1996,xi_ising_2016} (see Supplementary Information).
In Figure~\ref{fig-alldata}, we use this pair-breaking equation to fit the 1L~\tas and to the 1L~\nbse data \cite{xi_ising_2016}.
From the fits, we find that $\Hso^{\tas}\approx\SI{815}{T}$ and $\Hso^{\nbse}\approx\SI{340}{T}$, both of which are significantly larger than the largest in-plane magnetic fields we could apply (about \SI{34.5}{T}).
The fit to the 1L~\tas data in particular extrapolates to over $10\times$ the Pauli limit, or $\Hc(0)\approx\SI{55}{T}$, a larger upper critical field than any other thickness of \tas or \nbse.

Close to $\Tcz$, the pair-breaking equation can be approximated as
\begin{equation}
\Hc(T) \approx \sqrt{ \Hso \Hp \left( 1-\frac{T}{\Tcz} \right) },
\label{eq-hc2}
\end{equation}
which reflects the usual square-root dependence of two-dimensional Ginzburg-Landau theory  \cite{tinkham_introduction_1996}.
For the two monolayer compounds, which have nearly equal $\Tcz \approx \SI{3}{K}$ and therefore nearly equal $\Hp \approx \SI{5.5}{T}$, Eq.~\ref{eq-hc2} implies that the ratio of $\Hc(T)$ for \tas to that of \nbse should be close to $H_{c2}^{\parallel(\tas)}/H_{c2}^{\parallel(\nbse)}=\sqrt{\Hso^{\tas}/\Hso^{\nbse}} = \sqrt{\Dso^{\tas}/\Dso^{\nbse}}$, where $\Dso \propto \Hso$ is the spin-orbit splitting between the spin-up and spin-down bands at the Fermi level, estimated by averaging the $\mathbf{k}$-dependent spin-orbit splitting $\Dvb(\mathbf{k})$ over the Fermi surface, $\Dso \equiv \langle\Dvb(\mathbf{k}_F) \rangle_\mathbf{k}$.
A simple estimate from the DFT bands presented in Fig.~\ref{fig-bands}, considering only the \K,~\Kp pockets, yields $\Dso^{\mathrm{K},\tas} \approx \SI{236}{meV}$ and $\Dso^{\mathrm{K},\nbse} \approx \SI{104}{meV}$, which gives a ratio of $\Dso^{\tas}/\Dso^{\nbse} \approx 2.3$, in reasonable agreement with the experimentally determined ratio $\Hso^{\tas}/\Hso^{\nbse} \approx 2.4$.

However, the value of $\Hso=\SI{815}{T}$ determined from our fit for monolayer \tas suggests a spin-orbit splitting at the Fermi level of $\Dso^{\tas} = 2\muB\Hso^{\tas}=\SI{94.4}{meV}$, implying an overestimate by the DFT-computed splitting around K by $2.5\times$ the measured valued of \SI{94.4}{meV}.
From the band structure calculations shown in Fig.~\ref{fig-bands}a and \ref{fig-bands}c, it is clear that in \tas, there is also a significant effect of spin-orbit splitting of the Fermi surface pocket centered on the $\Gamma$-point, though the magnitude is smaller than around the \K and \Kp points (see Fig.~\ref{fig-alldata}c).
Including the $\Gamma$~pocket (i.e. using the full Brillouin zone) in the estimate of $\Dso^{\mathrm{BZ},\tas} = \SI{156}{meV}$ from the DFT bands reduces this disparity to $1.65\times$ the measured value.
We find that the band structure calculation for \nbse (Fig.~\ref{fig-bands}b and \ref{fig-bands}d), including the \K, \Kp, and $\Gamma$ pockets, similarly overestimates the strength of $\Dso^{\mathrm{BZ},\nbse}=\SI{62.6}{meV}$ compared to the measured value of \SI{39.5}{meV} (equivalent to \SI{340}{T}).

The pair-breaking equation with $\alpha = \muB (\Hpara)^2/\Hso$, with $\Hso$ as a fitting parameter, does a reasonable job of fitting the measured $\Hc(T)$, but \Hso is nevertheless smaller than expected from our band structure calculations.
One effect that can reduce $\Hc$ is Rashba SOC due to an out-of-plane electric field.
The Rashba SOC tends to tilt the electron spins in-plane and therefore competes with the Ising SOC, an effect considered in the studies of electrolyte-gate-induced superconductivity in \ce{MoS2} \cite{lu_evidence_2015,saito_superconductivity_2016}.   
A model including the competing effects of Rashba and Ising SOC is employed in Ref.~\citenum{lu_evidence_2015} to fit $\Hc(T)$ data from \ce{MoS2}. One feature of this model is that $\Hc(T)$ has a maximum at a finite temperature $T > 0$, whereas the pair-breaking equation employed in Fig.~\ref{fig-alldata} increases monotonically as $T$ is lowered to zero.
Although the lowest (reduced) temperature we could reliably use to extract $\Hc$ for 1L~\tas was about $0.8\,\Tcz$ (beyond which the transition exceeded the maximum available field of \SI{34.5}{T}), we note that in the 3L--5L~\tas and 2L~\nbse samples we could attain much lower temperature scales.
From the few-layer data in Fig.~\ref{fig-alldata}a and Fig.~\ref{fig-alldata}b, we observe that $\Hc(T)$ does not appear to have a maximum at finite temperature but rather increases monotonically as $T$ is lowered, casting doubt on the applicability of this model for few-layer \tas and \nbse (see SI).

In bilayer crystals the broken inversion symmetry of the monolayer system is restored, with an inversion center appearing between the layers of the bilayer.
In trilayers, global inversion symmetry is broken again; restored in four-layer crystals, and so on.
One might thus expect oscillatory behavior in the strength of Ising superconductivity as a function of the number of layers, however this is not what we observe in few-layer samples.
Despite the restored global inversion symmetry in, for example, bilayer \nbse, the upper critical field remains much higher than the Pauli paramagnetic limit, approaching 29 T (3.5 times $\Hp$) as $T\rightarrow0$ (Figs.~\ref{fig-perpandparallelfield} and~\ref{fig-alldata}), though the enhancement of $\Hc$ above $\Hp$ is significantly less than for the monolayer samples.
The observation of $\Hc > \Hp$ also holds for the 3L, 4L, and 5L~\tas devices as well, with only a weak dependence on the thickness below 1L.

The lack of dependence on layer number parity can be understood in terms of weak coupling between the layers.
In the limit of zero interlayer coupling, each layer superconducts independently and the strength of Ising SOC is equivalent to the monolayer system (Fig.~\ref{fig-alldata}d).
With a small amount of tunneling, say between $d$~orbitals of \ce{Ta} atoms in neighboring layers, the single-particle states in each layer will overlap with states experiencing an opposite effective field, $-\Bsov$, due to the $\pi$-rotated crystal field in the surrounding layers.
The net effect is weaker Ising SOC and a reduced degree of upper critical field enhancement compared to the monolayer case, especially for 2L and 3L crystals, with diminishing changes for additional layers beyond that (until orbital effects begin to dominate).

To gauge the strength of this effect, we consider an interlayer hopping energy $t_\perp$ extracted from DFT bands for 2L, 3L, and 4L \tas and \nbse without including SOC (Fig.~\ref{fig-alldata}e).
We estimate $t_\perp$ from the average dispersion (splitting) in the out-of-plane direction along the Fermi surface (see Methods) and plot the ratio $t_\perp/\Dso$ in Fig.~\ref{fig-alldata}, defined to be zero for 1L.
We find relatively weak interlayer coupling for states near the Fermi level, $t_\perp\approx\SI{10}{meV}$ for \tas and \SI{20}{meV} for \nbse, especially compared to the Fermi surface average SOC of $\Dso=\SI{156}{meV}$ for \tas and \SI{62.6}{meV} for \nbse.
To highlight the trend between increasing $t_\perp$ and decreasing $\Hc$ as they vary with layer number $N$, we plot $\Hc(N)$ directly below, in Fig.~\ref{fig-alldata}f.
We use the reduced quantity $h_{c2}\equiv \Hc/\Hp$ evaluated at $T/\Tcz = 0.8$ for both \tas and \nbse, and we normalize $h_{c2}$ to its value for the 1L device, $h_{c2}^{\mathrm{(1L)}}$.
For multilayer devices, the value of $h_{c2}/h_{c2}^\mathrm{(1L)}$ is only weakly dependent on $N$, but diminishes more rapidly for \nbse than \tas as $N$ is increased, trending inversely with the ratio $t_\perp/\Dso$, which is larger and increases faster for \nbse compared to \tas.

In terms of the underlying crystal symmetries, we interpret the measured weak dependence of $\Hc(N)$ on $N > 1$ to reflect the \emph{staggered} non-centrosymmetric structure of 2H-\tas and \nbse, wherein the individual layers lack \emph{local} inversion symmetry despite globally possessing inversion centers between the layers.
This type of structure is also found in the layered superconductor \ce{SrPtAs}, which exhibits a similar enhancement of the paramagnetic limit despite having a global center of inversion \cite{youn_role_2012}.
In the case of \ce{SrPtAs}, the individual hexagonal As-Pt superconducting layers can be considered to have a local inversion symmetry breaking and therefore retain some of the physical properties associated with non-centrosymmetric superconductivity, such as the enhanced $\Hc$.

Finally, we comment on the striking difference in $\Tcz(N)$ between \tas and \nbse.
Upon reducing the thickness of \tas in the few-layer limit, we find that $\Tcz$ increases from the bulk value of $\approx\SI{800}{mK}$ up to \SI{3}{K} in 1L~\tas (Fig.~\ref{fig-zerofield}c), however the origin of this enhancement remains to be understood.
The reversed trend $\Delta\Tcz/\Delta N < 0$ observed in ultrathin \tas is unusual not only compared to \nbse, but also in the context of other two-dimensional and layered superconductors (and echoes the well-studied trend of $\Tcz$ in intercalated bulk \tas, see SI).
We suggest detailed studies of the layer dependence of potentially competing charge-density-wave (CDW) order as a route to understanding $\Tcz(N)$ of few-layer \tas~\cite{klemm_pristine_2015,sanders_crystalline_2016,freitas_strong_2016,pan_enhanced_2017}.
In particular, scanning tunneling spectroscopy of the superconducting and CDW gaps in the few-layer limit may also benefit this understanding greatly~\cite{galvis_zero-bias_2014,ugeda_characterization_2015,talantsev_origin_2017}.

In conclusion, we have shown that encapsulated \tas is stable in its monolayer 1H phase, and that its upper critical field $\Hc$ is larger than that of \nbse by an amount that is well predicted by a phenomenological pair-breaking model including the effect of intrinsic spin-orbit coupling.
Band structure calculations of the total spin-orbit coupling around the Fermi surface point to the importance of the pockets around the \K and \Kp points, as well as the $\Gamma$ point in the metallic TMD superconductors.
Because of the larger spin-orbit coupling in \tas relative to \nbse, many phenomena such as spin-triplet superconductivity and modulated superconductivity may be more readily studied in atomically-thin \tas, especially the monolayer, which even in absolute terms appears to possess the largest $\Hc$ recorded for any thickness of either material.

\section*{Methods}

\subsection*{Device fabrication}

We create devices from few-micron-sized flakes of \tas and \nbse exfoliated from bulk 2H$_a$-polytype crystals.
Because both compounds are susceptible to degradation in ambient conditions, we exfoliate the crystals inside of a nitrogen-filled glove box and encapsulate the exfoliated flakes with hexagonal boron nitride (\hBN) in the same environment.

To rule out a crystallographic phase change of the few-layer crystals upon exfoliate from the bulk 2H$_a$ form, we consider both the anisotrpy of the upper critical field and the polarization of the second harmonic generation.
For example, the enhancement of $\Hc$ in 1L~\tas over $10\times$ the Pauli field limit relies on the lack on inversion symmetry in the monolayer crystal, ruling out the 1T phase, which is fully centrosymmetric within each layer.
Second harmonic generation of all \tas devices in the study exhibits a six-fold rose pattern in the azimuthal angle, reflecting the underlying three-fold symmetry of the 1H~phase and ruling out monoclinic 1T$^\prime$ and orthorhombic T$_\mathrm{d}$ phases.

To make electrical contact to the crystals, we transfer few-layer graphite, which is similarly exfoliated from bulk, and overlap with part of the TMD crystal to create areas with an atomically-smooth interface for electrical contact.
The overlapping region of graphite/TMD (entirely encapsulated by \hBN on top) is then etched into separate channels to allow 4-terminal measurements of the superconducting TMD alone (see Fig.~\ref{fig-zerofield} insets).
The etched graphite leads extend beyond the \hBN encapsulating layer allowing \ce{Cr/Pd/Au} leads with top-contact to the graphite to be defined using standard electron-beam lithography techniques.

\subsection*{Magnetotransport measurements}
Magnetotransport measurements were made using standard low-frequency AC lock-in techniques with SR8x0 series lock-in amplifiers and a Keithley 2400 SourceMeter. The samples were measured in a dilution refrigerator to a minimum temperature of \SI{25}{mK} and maximum field of \SI{12}{T}, as well as at the National High Magnetic Field Lab in Tallahassee, Florida in a \ce{He}-3 refrigerator to a minimum temperature of \SI{300}{mK} and a maximum field of \SI{34.5}{T}.

In parallel field:
This measurement is obtained by first fixing the temperature and the magnetic field and then varying the angle between the magnetic field and the sample until a minimum in the resistance is achieved, allowing us to precisely locate the parallel configuration.
We then sweep the magnetic field at fixed temperature $T$ and extract $\Hc(T)$ as the value of the field for which $R=\Rn/2$, as in Fig.~\ref{fig-perpandparallelfield}e.
Measuring $R(T)$ at fixed parallel field and varying temperature is principally equivalent, however operationally we find that it is more difficult to hold a perfect parallel position while also varying temperature, and thus we prefer the consistency of varying the field while holding the temperature fixed.
Nevertheless, we do take data in both modes and plot both data sets together, for example, as shown for the 5L \tas device in Fig.~\ref{fig-alldata}a, with squares coming from field sweeps and crosses coming from temperature sweeps.

\subsection*{Density Functional Theory calculations}

The calculations for the first-principles part was performed using the projector augmented wave~\cite{blochl_projector_1994,kresse_ultrasoft_1999,kresse_efficient_1996} method encoded in Vienna \textit{ab~initio} simulation package (VASP)~\cite{kresse_efficient_1996} with the generalized gradient approximation in the parameterization of Perdew, Burke, and Enzerhof~\cite{perdew_generalized_1996,perdew_generalized_1997}.
An outer shell configuration of  4p$^6$ 4d$^4$ 5s$^1$, 5d$^4$ 6s$^1$, [Ne] 3s$^2$ 3p$^4$ and [Ar] 4s$^2$ 4p$^4$  were used for Nb, Ta, S and Se respectively.
Structural optimization was performed for monolayers with a vacuum region more than \SI{15}{\angstrom}.
All the ions were relaxed so that the total energies converged to \SI{0.5}{meV/atom} with a regular $16 \times 16 \times 1$ Monkhorst-Pack grid.

After obtaining the \textit{ab~initio} wave functions from a self-consistent calculation the corresponding Fermi-surface was computed utilizing the Wannier interpolation approach~\cite{mostofi_wannier90:_2008,marzari_maximally_1997,souza_maximally_2001}.
When spin-orbit coupling was included, the spin degeneracy of the bands was lifted away from the $\Gamma$-point.
The Wannier interpolation was performed by projecting onto 22 bands at each $k$-point, 10 from the transition metal $d$~orbitals (spinors) and 12 from the two chalcogen atom $p$~orbitals.
The corresponding spin-projection along the Fermi-surface was obtained separately from the first-principles calculation using a Monkhorst-Pack grid of $108 \times 108 \times 1$, and was superimposed on the Fermi-surface obtained using the Wannier interpolation.

The interlayer coupling in multilayers were obtained by calculating the dispersion (total splitting) of the bands in the out-of-plane direction without including spin-orbit coupling~\cite{gong_magnetoelectric_2013}.
An interlayer coupling strength $t_\perp$ is then estimated by extracting half of the remaining splitting of the bands near the Fermi level, $t_\perp \equiv \Dvb(\mathbf{k}_F)/2$ without SOC.

\section*{Contributions}

SCB, MRS, DPG, and AWT fabricated the devices.
SCB, MRS, DPG, AWT, and BMH performed the magnetotransport experiments.
NS, with supervision from DX, performed the theoretical calculations.
KW and TT grew the hBN crystals.
KLS performed the optical characterization of the devices under the supervision of XX.
SCB and BMH wrote the paper, with contributions from MRS.    

\section*{Acknowledgements}
We acknowledge fabrication assistance from Jingyi Wu and helpful discussions with Y.~Yang, K.~F.~Mak, C.~R.~Dean, and E.~Telford.  We also thank D.~Graf and E.~S.~Choi for experimental assistance at the National High Magnetic Field Laboratory. Funding for SCB was provided by the Charles E.~Kaufman Foundation, a supporting organization of The Pittsburgh Foundation, via Young Investigator research grant KA2016-85226.  Work on device fabrication by DPG was supported by the Department of Energy Early Career program under award number DE-SC0018115.  Work on device fabrication and measurement by MRS was supported by the National Science Foundation PIRE program under award number 1743717.  The high-field measurements were made at the National High Magnetic Field Laboratory.
XX and DX are supported by the Department of Energy, Basic Energy Sciences, Materials Sciences and Engineering Division (DE-SC0012509).
NS acknowledges National Science Foundation [Platform for the Accelerated Realization, Analysis, and Discovery of Interface Materials (PARADIM)] under Cooperative Agreement No. DMR-1539918 and Cornell University Center for Advanced Computing for his time at Cornell University.

\bibliography{all_ising}

\onecolumngrid 
\cleardoublepage
\setcounter{figure}{0}
\setcounter{equation}{0}
\renewcommand\thesection{S\arabic{section}}
\renewcommand\thesubsection{S\thesection.\arabic{subsection}}
\renewcommand{\thefigure}{S\arabic{figure}}
\renewcommand{\theequation}{S\arabic{equation}}

\section{Supplementary Information}

\subsection{Fitting to pair-breaking equation}

In the presence of strong perturbations that break time-reversal symmetry, the destruction of superconductivity can be described in terms of Abrikosov--Gor'kov (AG) theory, often applied to the cases of magnetic impurity scattering, spin-orbit scattering, and thin films in magnetic fields (in the ``dirty'' limit); related formally by equivalence theorems~\cite{maki_gapless_1969}.
For second-order phase transitions from the superconducting to the normal state, self-consistent solution of the linear gap equation in AG theory leads to a universal pair-breaking equation,
\begin{equation}
\ln \left( \frac{\Tc}{\Tcz} \right) + \psi \left(  \frac{1}{2} + \frac{\alpha}{2 \pi \kB \Tc} \right) - \psi \left(  \frac{1}{2} \right) = 0,
\label{eq-pairbreak}
\end{equation}
which implicitly gives the transition temperature $\Tc$ as a function of a pair-breaking strength $\alpha$ that depends on the system ($\psi$ is the digamma function)~\cite{maki_gapless_1969,klemm_theory_1975,tinkham_introduction_1996}.
In our system the external perturbation is the applied in-plane magnetic field, modified by the presence of strong Ising SOC (which by itself does \emph{not} break time-reversal symmetry).
Due to the atomic thickness of our samples, orbital effects related to in-plane fields are quenched and thus the relevant pair-breaking mechanisms only involve the spin interaction with the external field.

The spin part of the Hamiltonian has two terms,
\begin{equation}
H(\mathbf{k}) = \muB \tau \Bso(\mathbf{k}) \sigma_z + \muB \Bpara \sigma_x,
\label{eq-spinH}
\end{equation}
where $\Bso$ gives the momentum-dependent magnitude of the effective spin-orbit field ($\equiv |\Bsov|$ in main text), $\tau=\pm 1$ labels the valley pseudospin, and $\sigma_i$ are Pauli matrices which operate in real spin space.
The eigenvalues for a given $\mathbf{k}$ (dropping the explicit argument) are
\begin{equation}
E_\pm = \pm \muB \sqrt{\Bso^2 + \Bpara^2} \approx \pm \muB \left[ \Bso + \Bpara^2/2\Bso + \cdots \right]
\end{equation}
where the term quadratic in $B_\parallel$ is the lowest order energy that contributes to pair-breaking, since the intrinsic SOC term does not break time-reversal symmetry.
Using the total Zeeman splitting caused by the introduction of the in-plane field, we invoke Eq.~\ref{eq-pairbreak} with $\alpha = \muB \Bpara^2/\Bso = \mu_0 \muB \Hpara^2/\Hso$ with $\Hso$ as a free parameter to fit our monolayer $\Hc(T)$ data.

For small $\alpha$, Eq.~\ref{eq-pairbreak} can be expanded around $1/2$ to give
\begin{equation}
\kB \left( \Tcz - T \right) = \frac{\pi}{4} \alpha
\end{equation}
which yields (for $T \rightarrow \Tcz$)
\begin{equation}
\Hc = \sqrt{ \frac{4\Hso}{\pi\muB} \kB \Tcz \left( 1 - \frac{T}{\Tcz} \right) }.
\end{equation}
Substitution of the Pauli field $\Hp = \Delta_0/\sqrt{2}\muB = 1.76 \kB \Tcz / \sqrt{2}\muB$ yields Eq.~\ref{eq-hc2}, with a prefactor equal to $2^{5/4}/\sqrt{1.76\pi}=1.01$.

\newpage
\subsection*{Fitting to phenomenological tight-binding model}

To explore the possibility of a competing Rashba effect in our TMD superconductors, we use the results of a phenomenological model employed by Refs.~\citenum{lu_evidence_2015} and \citenum{saito_superconductivity_2016} to describe $\Hc$ in monolayer \ce{MoS2}.
The starting point of this model is a normal-state Hamiltonian that includes both intrinsic (Ising) SOC as well as Rashba SOC:
\begin{equation}
H(\mathbf{k}+\tau\mathbf{K}) = \epsilon_\mathbf{k} + \frac{\Dso}{2}\tau\sigma_z + \alphaR(\sigma_y k_x - \sigma_x k_y) + \muB \mathbf{B} \cdot \boldsymbol{\sigma}
\label{eq-HwithRashba}
\end{equation}
where $\epsilon_\mathbf{k} = \hbar^2 k^2/2m -\mu$, a parabolic band approximation, $\tau=\pm 1$ is a valley index, $\Dso$ is the magnitude of the spin-splitting around the Fermi surface as defined in the main text, $\alphaR$ is the strength of Rashba SOC, and $\boldsymbol{\sigma}$ are the Pauli spin matrices.
The authors then derive an expression for a linearized gap equation that relates $\Hc$ and $T$, which is described by the two parameters $\Dso$ and $\alphaR$.
We have used this expression to fit the monolayer \tas and \nbse data (Fig.~\ref{fig-alldata}).

With $\alphaR=0$, this model estimates $\Dso=\SI{13}{meV}$, a significantly smaller value than the best fit of \SI{94.4}{meV} from the pair-breaking equation (details of this fitting are in the SI) and the extracted value \SI{156}{meV} from first-principles.
Thus, we take $\Dso=\SI{94.4}{meV}$ from the optimal fit of the pair-breaking equation to estimate the strength of $\alphaR$ necessary to fit our monolayer \tas data in the available temperature range.
We find that this phenomenological model requires a relatively large Rashba SOC energy of $\alphaR \kF=\SI{5.9}{meV}$ in order to describe our data.

	\begin{figure}[hb!]
	\begin{center}
\includegraphics[width=80 mm]{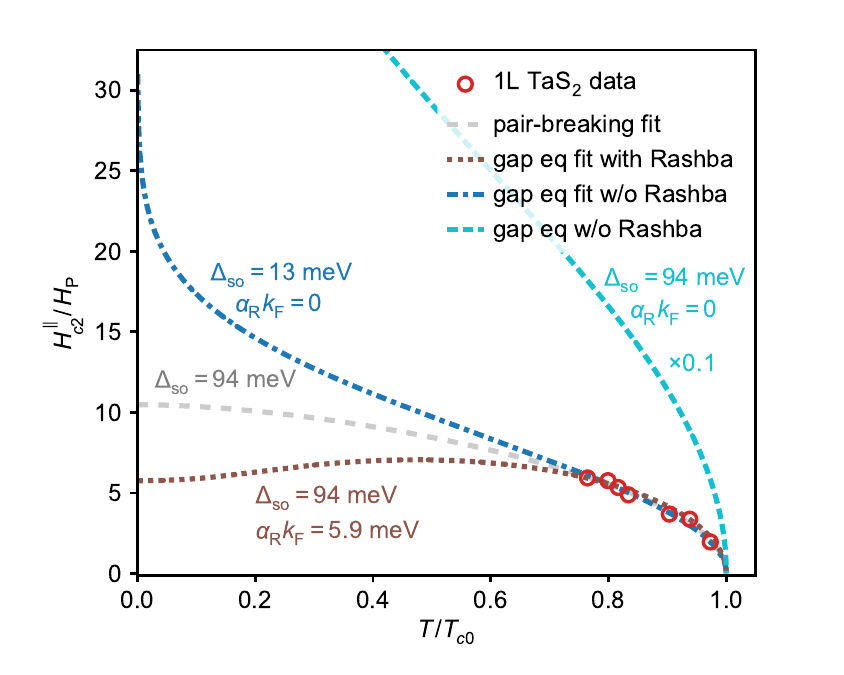}
\caption{
\textbf{Theoretical fitting of measured 1L~\tas parallel-field data.}
Best-fit curves to 1L~\tas data using single-parameter pair-breaking fit (Eq.~\ref{eq-pairbreak}; gray dashed), model including SOC and Rashba SOC (Eq.~\ref{eq-HwithRashba}; dotted) with $\Dso$ fixed to pair-breaking value, and the latter model with Rashba fixed at zero (dash-dot), allowing $\Dso$ to vary.
A curve from the same model, but using best-fit pair-breaking value $\Dso=\SI{94.4}{meV}$ without Rashba (cyan dashed; scaled $0.1\times$ for plotting) is also included to highlight sensitivity of the model to the value of $\alphaR \kF$.
}
		\label{fig-microscopic}
	\end{center}
\end{figure}

\newpage

\subsection{Background subtraction on three-terminal devices}

\begin{figure}[hb!]
	\begin{center}
\includegraphics[width=178 mm]{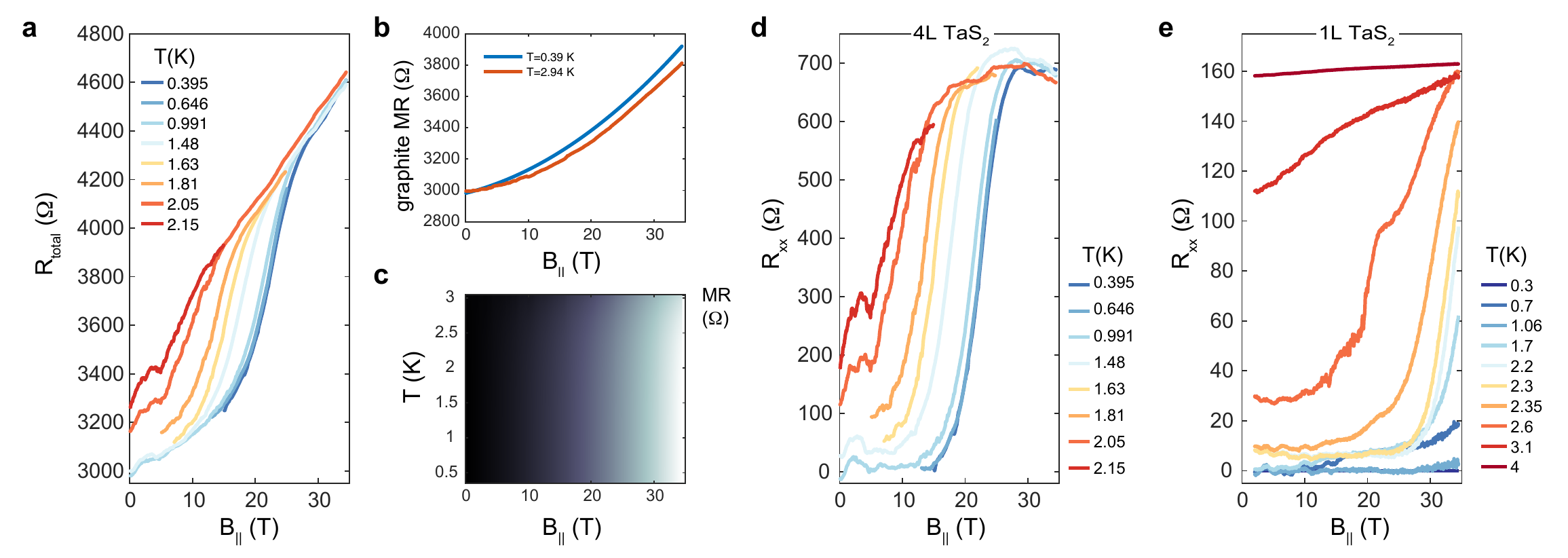}
\caption{
\textbf{Procedure for background subtraction of graphite magnetoresistance.}
(a) Three-terminal measurement of $R_\mathrm{total}$ for the 4L~\tas device, which includes a background magnetoresistance $\Rgr$ from the graphite contacts.
(b) High-$T$ (\SI{2.94}{K}) and low-$T$ (\SI{0.39}{K}) curves used to generate background $\Rgr(\Bpara,T)$.  The high-$T$ curve (2.94 K) has the normal state resistance $R_n\approx 700\Omega$ subtracted; the low-$T$ curve (0.395 K) was generated by fitting a polynomial to the 0.395 K data from (a) for $B<\SI{12}{T}$.
(b) Surface of $\Rgr(\Bpara,T)$ generated by interpolating between the two curves from (b).
(d) Subtracted $R(\Bpara)$ curves at various $T$ for 4L~\tas, which are used to extract $\Hc(T)$ as shown in Fig.~\ref{fig-alldata} of the main text. Temperature legend is same as in (a).
(e) Subtracted $R(\Bpara)$ curves at various $T$ for 1L~\tas, generated by an identical subtraction procedure as described for the 4L device. }
		\label{fig-graphiteMR}
	\end{center}
\end{figure}

In two of our devices (1L and 4L~\tas), broken contacts forced us to measure $R_{\mathrm{total}}(\Bpara,T)$ in a three-terminal geometry (e.g. Fig.~\ref{fig-graphiteMR}a). This meant that we could not eliminate one of the contact resistances from our measurement, each contact consisting of graphite in series with Cr/Pd/Au (see Methods).  Because of the temperature-dependent, strong magnetoresistance (MR) of graphite, the procedure used to subtract the background was non-trivial.  We describe it here.

Without a direct measurement of the MR of graphite as a function of $\Bpara$ and $T$, we generate a surface $\Rgr(\Bpara,T)$ by interpolating between two curves: (i)  $R(\Bpara,T>T_c)$, where the total resistance is the graphite MR at $T>T_c$ plus a constant offset ($\Rn$ of the superconductor in the normal state, with negligible MR), and (ii) $R(\Bpara,0)$, where  the device is superconducting for low fields. For the latter, a reasonable guess can be generated for $\Rgr(\Bpara,0)$ by fitting the low-field portion of $R_{\textrm{total}}(\Bpara,0)$ to a polynomial, or in the case of the 1L~\tas data, by simply using the $R_{\textrm{total}}(\Bpara,0)$ curve as $\Rgr(\Bpara,0)$ since the device does not go normal for any field in this limit.

Since the background curves are relatively insensitive to temperature (Fig.~\ref{fig-graphiteMR}b and \ref{fig-graphiteMR}c), the values of $\Hc(T)$ extracted from this procedure are quite insensitive to precisely how the interpolation was performed.  For example, we can estimate that difference in the value of $\Hc(T)$ extracted directly from the raw data differs from that extracted by a linear interpolation of $\Rgr(\Bpara,0)$ by at most $\pm 1$ T. A similar conclusion can be drawn for the difference in $\Hc(T)$ deduced from a linear interpolation versus an exponential interpolation of $\Rgr(\Bpara,0)$.

\newpage

\subsection{Control experiments on bulk devices}

\begin{figure}[hb!]
	\begin{center}
\includegraphics[width=89 mm]{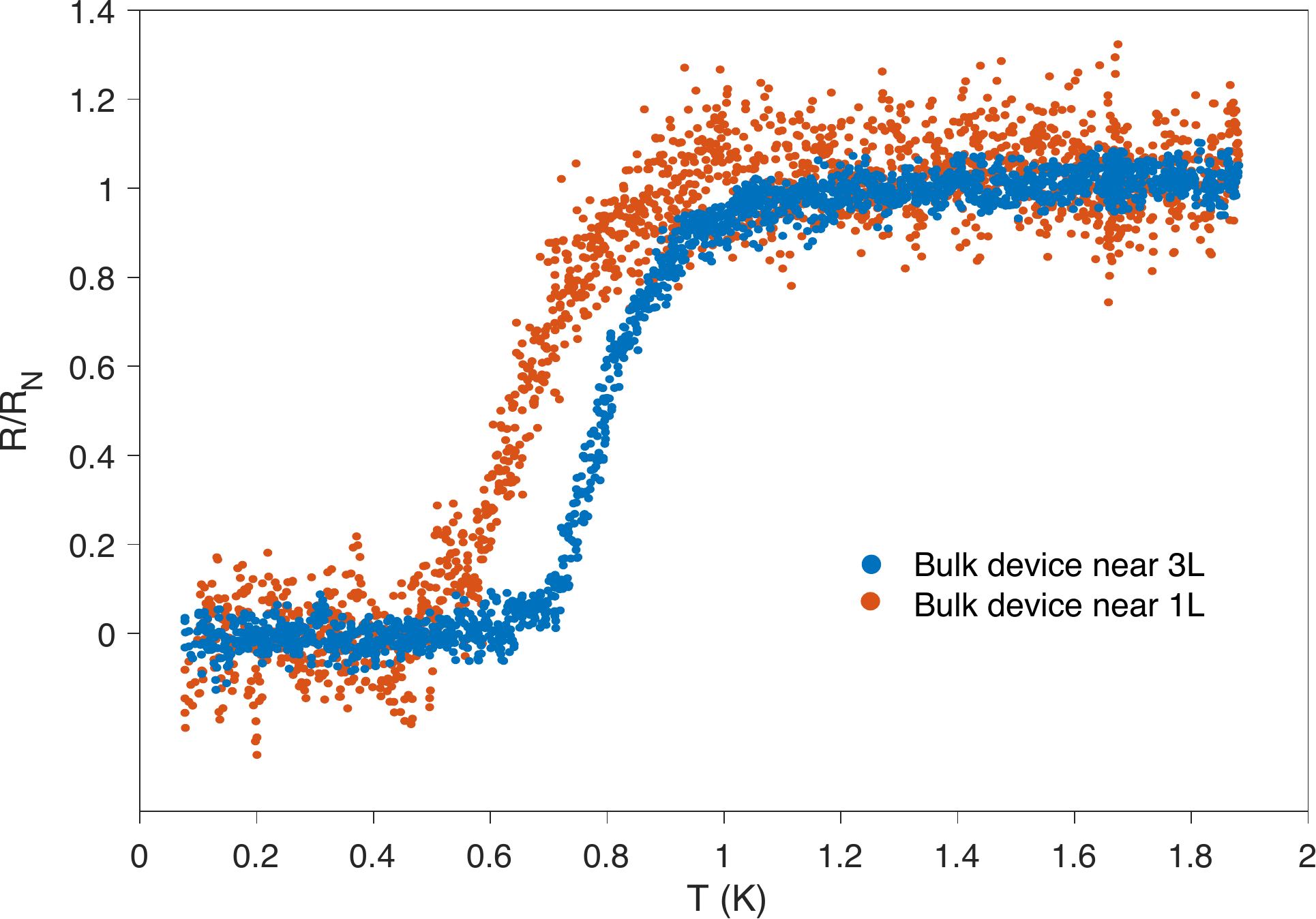}
\caption{
\textbf{Confirmation of bulk \tas properties.}
Temperature dependence of bulk \tas sheet resistance for two-terminal devices fabricated using flakes $\gg\SI{20}{nm}$ thick, nearby the 1L and 3L few-layer devices.
}
		\label{fig-bulk}
	\end{center}
\end{figure}

In some devices that we studied, the enhancement of $\Tc$ above the bulk value may be due to another mechanism besides the reduction in layer number $N$. For example, it is well known that intercalation of \tas with certain organic molecules and elemental metals can both increase $\Tc$ and enhance $\Hc$ \cite{klemm_pristine_2015}.  To exclude this possibility for the $\Hc$ data presented in Fig.~\ref{fig-alldata} for the 1L and 3L samples, we fabricated quasi-bulk devices from thicker flakes from the same batch of exfoliation (and on the same chip) as these two thin samples as two control devices (Fig.~\ref{fig-bulk}).  The control devices were therefore exposed to the identical fabrication processes as the thin devices, with the exception of encapsulation by \hBN/graphite contacts.  The two bulk devices exhibited much lower $\Tc$ than the thin devices on the same flake that were comparable to the expected value of the bulk $\Tc=$~\SIrange{0.5}{0.8}{K}, which provides evidence that the enhancement of $\Tc$ in the thin devices (and therefore also the enhancement of $\Hc$) is indeed due to the reduction of $N$.  An additional piece of evidence supporting this idea is the discrepancy between the shape of the $\Hc(T)$ curves for intercalated \tas, which have a finite slope $d\Hc/dT$ and upward curvature at $\Tcz$, and are generally well-fit by the Klemm-Luther-Beasley theory \cite{klemm_theory_1975}.  Our monolayer and few-layer \tas devices exhibit very steep (infinite) derivative at $\Tcz$, showing downward curvature and no inflection point as $T$ is lowered. 

\begin{figure*}[ht!]
	\begin{center}
\includegraphics[width=178mm]{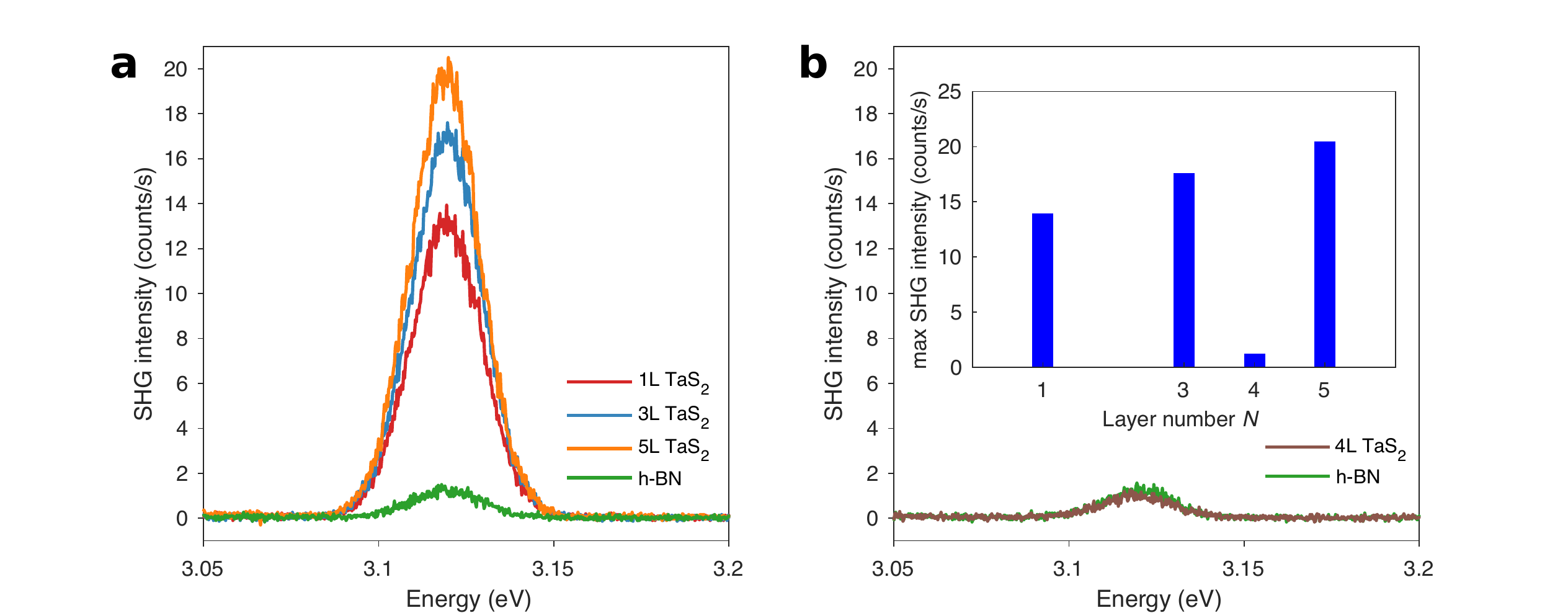}
\caption{ 
\textbf{Second harmonic generation of \tas devices.}
(a) SHG spectra for three odd layer number \tas devices with h-BN spectrum for comparison.
(b) SHG spectra for an even layer number \tas device with h-BN spectrum for comparison.
Inset: maximum SHG intensity observed for the four samples of various thickness.
}
		\label{fig-SHG}
	\end{center}
\end{figure*}

\subsection{Second-harmonic generation study of \tas devices}

Second-harmonic generation (SHG) is a sensitive probe of inversion symmetry breaking, since the second-order nonlinear susceptibility is only nonzero for noncentrosymmetric crystals \cite{boyd_nonlinear_2003}.
In few-layer 2H-phase TMDs, strong SHG is only observed in samples with odd layer number, where inversion symmetry is broken \cite{li_probing_2013}.
We studied the SHG from our \tas samples to clearly distinguish between even and odd layer number, and thus further corroborate our sample thickness assignments.
The samples were excited by a mode-locked Ti:sapphire laser (76 MHz repetition rate, 200 fs pulse duration, 1 mW average power) at 795 nm under normal incidence in reflection geometry and the SHG was detected by a spectrometer and Si charge-coupled device.
The measurements were performed at room temperature under vacuum, and laser intensity was carefully tuned to prevent damage to the TMD material.
In Fig.~\ref{fig-SHG}, we show the SHG spectra from \tas samples.
The spectra of three odd layer number samples (1L, 3L, and 5L) are shown in Fig.~\ref{fig-SHG}a, while the spectrum of an even layer number (4L) sample is shown in Fig.~\ref{fig-SHG}b.
It can clearly be seen that the SHG intensity of the odd layer number \tas samples is significantly greater than that of the thin bulk h-BN that encapsulates them.
On the other hand, the even layer number \tas sample produces an intensity nearly equivalent to that of the encapsulating h-BN layer, showing that the 4L \tas does not contribute significantly to the SHG response.
The inset of Fig.~\ref{fig-SHG}b shows the maximum intensity observed for each of the four samples of varying thickness, demonstrating an oscillation in maximum intensity between odd and even layer number samples, similar to other 2H-phase TMDs \cite{li_probing_2013}.
Although a linear increasing trend is observed among the odd layer number samples, we draw no conclusion regarding this trend because of the small sample size.
The even and odd layer number assignments established by SHG are consistent with the thickness determined by transport and AFM.
 
\subsection{Potential for spin-texture tunability in the Fermi surface of monolayer \tas}

\begin{figure}[ht!]
  \begin{center}
    \includegraphics[width=89 mm]{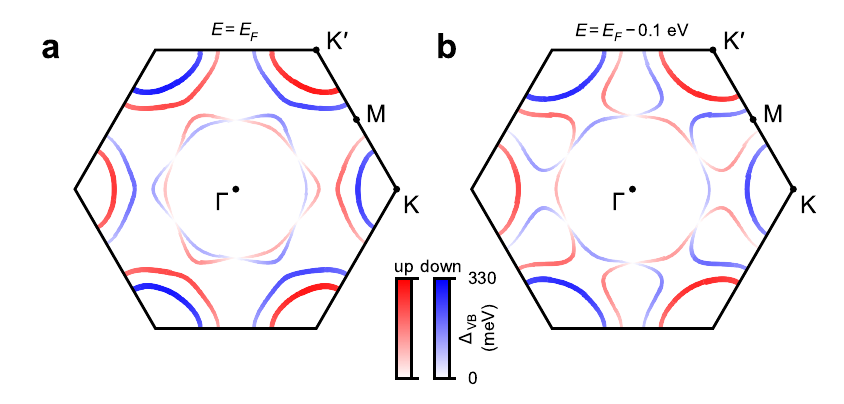}
\caption{\textbf{Qualitative change in \tas Fermi surface upon lowering the Fermi level.}
Fermi surface of monolayer \tas computed by density functional theory using (a) the intrinsic Fermi level and (b) an adjusted Fermi level \SI{0.1}{eV} lower than the intrinsic case, below the saddle-point along $\Gamma$--K of the upper spin band.
}
		\label{fig-newfs}
	\end{center}
\end{figure}

The intrinsic Fermi level of 1H-\tas as computed by DFT is within \SI{32}{meV} of a saddle point along $\Gamma$--K in the upper spin band, as shown in Fig.~\ref{fig-bands}a.
Two additional saddle points occur in a larger energy window: one from the lower spin band at the same location in $\mathbf{k}$-space and one at the spin-degenerate M~point, \SI{234}{meV} and \SI{278}{meV} below the intrinsic Fermi level, respectively.
Each of these saddle points is associated with a divergence in the density of states resulting from a van Hove singularity.
Tuning the Fermi level in a monolayer of \tas to coincide with any of these saddle points might therefore be expected to produce interesting consequences for the superconductivity, for example, producing additional enhancement in the $\Tcz$ due to the divergent density of states.
Such tuning may be possible via doping or electrostatic gating techniques as employed in other TMD superconducting materials \cite{ye_superconducting_2012,lu_evidence_2015,li_controlling_2016,saito_superconductivity_2016,xi_gate_2016,lu_full_2017}.
Furthermore, tuning the Fermi level to lie between the two saddle points along $\Gamma$--K would lead to a complete spin-polarization of the states along the $\Gamma$--K line of symmetry within each irreducible wedge of the Brillouin zone (Fig.~\ref{fig-newfs}).
As shown in Fig.~\ref{fig-newfs}b, lowering the Fermi level by as little as \SI{100}{meV} would be sufficient to qualitatively alter the Fermi surface geometry from the intrinsic case.
Such a spin configuration in momentum space may be favorable for unconventional superconducting pairings such as chiral or spatially-modulated states \cite{hsu_topological_2017}.

\subsection{Additional comments on layer dependence of $T_c$}

The critical temperature in bulk \tas is known to increase from the intrinsic value $\Tcz\approx\SI{0.8}{K}$ to as high as \SI{6}{K} when intercalated with various organic molecules, experiencing an associated increase in interlayer spacing and exhibiting anisotropic critical field behavior \cite{coleman_dimensional_1983} indicative of a dimensional crossover from 3D to 2D, as described by the theory of Klemm, Luther, and Beasley \cite{klemm_theory_1975}.
The exact origin of this enhancement is not fully understood, but the suppression of competing charge-density-wave (CDW) order has been suggested to contribute in intercalated \tas and other TMD layered compounds \cite{klemm_pristine_2015}.
Evidently, a $\Tcz$ enhancement up to \SI{3}{K} was also observed in randomly-stacked \tas nanosheets with a random twist angle between the layers, and no CDW transition was observed in the twisted stacks from \SIrange{2}{300}{K} \cite{pan_enhanced_2017}.
Epitaxial 1H-\tas formed on \ce{Au}(111) substrates also does not exhibit a CDW when measured directly by scanning tunneling microscopy at \SI{4.7}{K} \cite{sanders_crystalline_2016}, although the authors note that the most likely explanation is possible interface doping from the \ce{Au} substrate.
On the other hand, an unexpected \emph{coexisting} CDW phase was recently shown to persist along with enhanced superconductivity in bulk \tas under increased pressures up to at least \SI{25}{GPa}, with an associated increase in $\Tcz$ up to \SI{8.5}{K} at \SI{9.5}{GPa} \cite{freitas_strong_2016}, which seems to suggest that competition between CDW and superconducting order in these materials may not be universally applicable.
In another proposal, the enhancement of $\Tcz$ in \tas thinned down to \SI{3.5}{nm} is suggested to result from a strong, repulsive Coulomb interaction that is renormalized as the number of layers is reduced \cite{navarro-moratalla_enhanced_2016}.
Finally, in a recent theoretical study of several thin-film type-II superconductors with thicknesses less than the out-of-plane coherence length, a rather general model for the opening of a second superconducting gap that coexists with the bulk gap was linked to an enhancement of $\Tcz$ \cite{talantsev_origin_2017}.
Although we find that the increasing trend in atomically thin \tas continues to the monolayer limit, we cannot distinguish between the aforementioned mechanisms based on our magnetotransport experiments.

\end{document}